# A Novel GDP Prediction Technique based on Transfer Learning using $CO_2$ Emission Dataset

**Sandeep Kumar   Pranab K. Muhuri**
Department of Computer Science
South Asian University, Akbar Bhavan, Chanakyapuri
New Delhi-110021, India
Email: sandeepkumar@students.sau.ac.in;   pranabmuhuri@cs.sau.ac.in

*Abstract* - In the last 150 years, $CO_2$ concentration in the atmosphere has increased from 280 parts per million to 400 parts per million. This has caused an increase in the average global temperatures by nearly 0.7 °C due to the greenhouse effect. However, the most prosperous states are the highest emitters of greenhouse gases (especially $CO_2$). This indicates a strong relationship between gaseous emissions and the gross domestic product (GDP) of the states. Such a relationship is highly volatile and nonlinear due to its dependence on the technological advancements and constantly changing domestic and international regulatory policies and relations. To analyse such vastly nonlinear relationships, soft computing techniques has been quite effective as they can predict a compact solution for multi-variable parameters without any explicit insight into the internal system functionalities. This paper reports a novel transfer learning based approach for GDP prediction, which we have termed as 'Domain Adapted Transfer Learning for GDP Prediction. In the proposed approach per capita GDP of different nations is predicted using their $CO_2$ emissions via a model trained on the data of any developed or developing economy. Results are comparatively presented considering three well-known regression methods such as Generalized Regression Neural Network, Extreme Learning Machine and Support Vector Regression. Then the proposed approach is used to reliably estimate the missing per capita GDP of some of the war-torn and isolated countries.

## 1. Introduction

The rise in global warming due to greenhouse gases is unleashing catastrophic alterations in the climate. With nearly 99.4% parts per billion (ppb), $CO_2$ is the largest greenhouse gas in the atmosphere [19]. Its contribution in the global warming (excluding water vapour) is more than 72% (see Fig. 1). During the last 150 years $CO_2$ concentration in the atmosphere has increased from 280 parts per million to 400 parts per million. This has pushed up average global temperatures by nearly 0.70 °C over the last century [41].

The Paris Agreement, which came into force on 4 November 2016, and has been ratified by 185 Parties out of 197 signatories to the convention, aimed to not allow the rise in temperature in this century to be more than 2 °C above pre-industrial levels or to even push it further down to 1.5 °C. These 179 parties are responsible for more than 55% of the total global greenhouse gas emissions, a predefined threshold required for the ratification of the Paris Agreement [42], [48]. The mandate hopes for rapid reduction of global greenhouse gas emissions to ensure minimum impact and risk that may arise from the climate change. Satellites are the most effective tools in tracing the source of greenhouse gas emissions as well as to oversee if countries are abiding by their commitment to the Paris agreement. So far, only Japan and the United States of America (USA) had satellites that could monitor greenhouse gases emissions. On 22 Dec 2016, China has also launched a carbon dioxide monitoring satellite named 'TanSat' [41]. Such satellites monitor the distribution, concentration, and flow of carbon dioxide ($CO_2$) in the atmosphere. Also, first-hand emission data obtained by these satellites has been quite useful for researchers worldwide for analysing climate changes and its effect on the ecosystem [13], [17], the spending strength of its people [26], estimation of $CO_2$ sinks [11], and ocean acidification [35], etc.

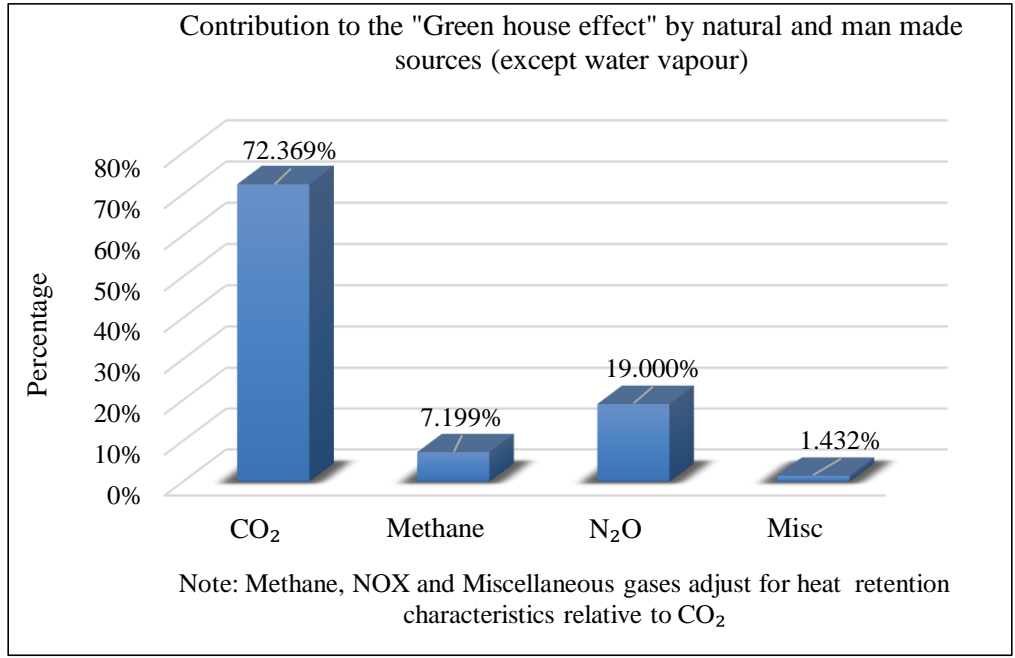

Fig. 1: Green House effect (except water vapour) [19].

According to Article 2 of the 'Paris Agreement', developing economies are being given more time [28] to achieve the desired levels of greenhouse emissions, so that their development needs are not compromised. Since, most devices used in industrialization (vehicles [45], electrical machinery and factories [46] etc.) draw their energy from non-renewable gaseous/ liquid/ solid fuels or renewable biofuels [61]. $CO_2$ is the most common by-product of these fuels. Hence, the fuel consumption trend of a nation can effectively predict its GDP [26]. The GDP estimation of a country can indicate its living standard and help in developing better economic policies. The fuel consumption of a country can easily be deduced by monitoring its fuel import, and also by analysing its satellite imagery to estimate its domestic fuel production and its $CO_2$ emissions.

This paper proposes a novel methodology for reliable estimation of GDP using the $CO_2$ emission data. Our approach is quite effective in predicting the GDP of a nation even when its domestic macroeconomic data is unavailable, insufficient or not authentic. The proposed method is based on a new technique called 'Transfer Learning (TL)', which has emerged as an attractive area of research in the last few years. The inspiration for TL is derived from human behaviour. Humans are the most advanced species and easily adapt to dynamic environments. It is also a human skill to repurpose and reuse information during the learning stage [38], which was the foundation for the evolution of TL. The difference between TL and traditional machine learning is depicted in Fig. 2. It pictorially shows how machine learning (ML) based models are built from scratch for every new task [27], while TL, like humans, utilizes previously learnt experience to solve a different but related task. TL has been useful in various real-life applications such as failure prediction [67], image segmentation [68], soil spectroscopy [69], Deep learning based classification [70], text-to-speech synthesis [71], and multi-objective optimization [72].

We have experimentally investigated the proposed approach considering three well-known regression methods such as support vector regression (SVR), generalized regression neural network (GRNN) and extreme learning machine (ELM) and presented the results in a comparative fashion. To train the proposed model, the $CO_2$ emission data and GDP data of a nation is used as input and output respectively. This trained model is then applied to predict the GDP of a geographically and economical dissimilar country, as compared to the one used for training. These experiments are first validated interchangeably for developed countries, developing country and least developed country. It is well known that these countries are highly diverse in terms of development, geographical location, education, etc. Since there are huge differences in the data distribution among the source domains and target domains, these experiments conform to the principles of transfer learning.

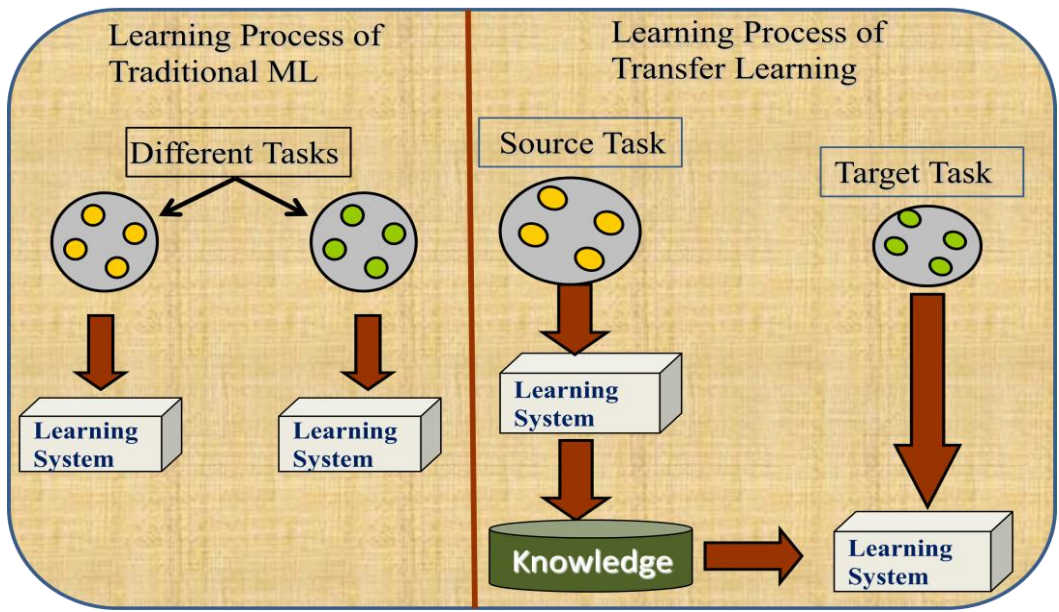

Fig. 2: Machine learning *vs*. Transfer learning [32]

Also, there are a number of countries, for which the GDP values are unavailable for a variety of reasons [51]. Few of such countries[1] are: Poland, Switzerland, Iraq, Syria, Myanmar, Yemen, and Afghanistan. For these countries, there is no available technique which can reliably predict their per capita GDP values for the missing periods. In this paper, we have demonstrated that the proposed approach is quite capable in predicting the missing per capita GDP of these countries.

The major contributions of the paper are summarized as follows:

1. This paper proposes a novel transfer learning based approach, which we termed as 'Domain Adapted Transfer Learning for GDP prediction' (DATL-GDP), for precise prediction of the per capita GDP values of a country using its $CO_2$ emission data.

2. It demonstrates the working of the proposed approach considering the $CO_2$ emission data belonging to four different nations with heterogeneous economic background.

---

[1] Among these countries, GDP values of Poland and Switzerland are not available for the durations (1960-1989) and (1970-1979), respectively; Iraq, Syria, Yemen and Afghanistan are among the world's most war-torn countries for more than a decade or so; Myanmar had been isolated for the duration (1960-1999), due to which their macroeconomic data were not reliably available for accurate estimation of their GDP.

Specifically, it considers two developed economies (USA, EU), one developing economy (IND), and one highly indebted, poor or least developed economy (CMR).

3. It investigates three different regression methods (viz., SVR, GRNN, and ELM) to assess their comparative preciseness in GDP prediction using $CO_2$ emission data.

4. It empirically demonstrates the efficacy of the proposed approach in predicting the missing per capita GDP values of a diverse range of countries comprising isolated/war-torn countries.

5. It first validates the prediction capabilities of SVR, GRNN and ELM on a small portion of the available GDP values of a country, then the best among them is used for precise estimation of the missing per capita GDP values of that country.

6. Thus, the proposed DATL-GDP approach provides an alternative way to the existing methodologies for estimating the per capita GDP of a country by using its $CO_2$ emissions data, which is quite effective even when macroeconomic data of a country is unavailable.

The nomenclature of the symbols and the abbreviations used in this paper are given in the Table 1.

Table 1: Nomenclature of the symbols and the abbreviations

| Symbols/ Abbreviations | Description | Symbols/ Abbreviations | Description |
|---|---|---|---|
| $C$ | Regularization | N | Number of training data |
| CMR | Cameroon | P(X) | Marginal probability distribution |
| $CO_2$ | Carbon dioxide | pdf | Probability distribution function |
| $D_s$ | Source domain | ppb | Parts per billion |
| $D_t$ | Target domain | $R^2$ | Coefficient of determination |
| $D_m$ | Mixed domain | $R^m$ | $m$ dimensional feature space |
| DATL-GDP | Domain Adapted Transfer Learning for GDP Prediction | RMSE | Root mean square error |
| EKC | Environmental Kuznets curve | RRMSE | Relative RMSE |
| $\varepsilon$ | epsilon | SLFNs | Single hidden layer feed forward neural networks |
| EU | European Union | $\sigma$ | Spread parameter |
| ELM | Extreme Learning Machine | SVM | Support vector machine |
| $f(.)$ | Predictive function | SVR | Support Vector Regression |
| F | Feature Space | TL | Transfer learning |
| $F_s$ | Source Domain feature space | USA | United States of America |
| $F_t$ | Target Domain feature space | $TD$ | Amount of target domain data used during training |
| FDI | Foreign direct investment | T | Learning task |
| $g()$ | Activation function | $T_s$ | Source learning task |
| GRNN | Generalized regression neural network | $T_t$ | Target learning task |
| GDP | Gross domestic product | $T_m$ | Mixed learning task |
| $H^{\hat{}}$ | Moore-Penrose generalized inverse | TL | Transfer learning |
| $h(x)$ | Output of hidden nodes | USA | United States of America |
| H | Hidden node output matrix | $\ddot{\omega}$ | Vector of 1's |
| IND | India | $\omega$ | Weight matrix of final layer |
| $K(x_1, x_j)$ | Kernel function | $(x_i, y_i)$ | $i^{th}$ training data point and its label |
| $K$ | Kernel matrix | Y | Label's of training data |
| ML | Machine learning | $\bar{Y}$ | Predicted labels |

Rest of the paper is organised as follows. Section 2 contains a literature survey. The technical terminology and the preliminaries are described in Section 3. Section 4 details the proposed approach. Experimental results and performance analysis are discussed in Section 5. Section 6 explains the benefits and application of the proposed methodology. Finally, Section 7 contains the conclusion deduced from the experiments and outlines the scope for future work.

## 2. Literature Survey

There are a number of $CO_2$ emission related studies in the literature. Hoa et al. studied the economic impact of $CO_2$ emissions and its mitigation policies in [20]. Narayan et al. [30] used cross-correlation estimation to analyse the association between economic growth and $CO_2$ emissions based on data from 181 countries. Abid [47] empirically analysed 41 EU countries and 58 Middle East & African countries to test the hypothesis of the Environmental Kuznets Curve (EKC). The experiments illustrated a monotonically increasing relationship between $CO_2$ emissions and GDP. EKC hypothesizes that various pollutants and per capita income follow an inverted-U-shaped relationship, i.e., in the early stages of economic development, when per capita income is low, the quality of the environment degrades significantly, but later, when per capita income increases the quality of the environment improves [52]. The EKC hypothesis was also supported by Yang et al. [73] in an analysis of relationship between per capita GDP and economy-related per capita greenhouse gas emissions of Russia. Fan et al. [14] conducted an excellent study focusing on the twenty different $CO_2$ utilization technologies of developing countries using data envelopment analysis based efficiency evaluation. They also suggested better and more effective decisions that governments and enterprises could take in order to use various emission reducing techniques, e.g., geological technologies, algae to food/feed additive technologies, etc. Renewable biofuels from biomass such as first generation biofuels [63], microalgal biofuels (algal oil) [64] or fuel pellets from woody biomass as substitutes to coal [65] etc. have proved to be effective replacements of conventional fuels, thus reducing $CO_2$ emission without stifling energy needs. The downside of these approaches is that they are costly, as in case of algal oil, which becomes effective only if crude petroleum sells for at least $100 per barrel [64]. In addition, renewable fuels from biomass can change land use and thereby increases food and fuel competition.

Govindaraju et al. [16] have suggested that the $CO_2$ emission of a nation is proportional to its economic might, which represents its level of industrialization. Arvin et al. [3] investigated the relationship between $CO_2$ emissions, urbanization and economic growth in the context of G-20 countries. Pao et al. [33] presented a case study of BRIC nations on the multivariate causal connection between $CO_2$ emissions, GDP, energy consumption, and FDI. The contrast between urban conglomerates [15] and rural folk [9] with regard to $CO_2$ emissions was also analysed with respect to china and for Asia in [24]. The importance of time dependent variations in transforming the relationship between energy consumption, $CO_2$ emissions and income was studied by Ajmi et al.[2]. For Gulf countries, Salahudin et al. [37] also analysed these specific casualties among energy consumption, $CO_2$ emissions, and income. Heidari et al. [18] reported that there exists a nonlinear relationship between energy consumption per capita, $CO_2$ emissions per capita, and gross domestic product (GDP) per capita. Begum et al. [4], [39] analysed the connection between economic growth, population growth, technological innovation, and trade openness and $CO_2$ emissions in the case of Malaysia. Saidi et al. [36] demonstrated the positive impact of energy consumption and $CO_2$ emissions on economic growth by analysing the data of 58 countries collected over the period (1990–2012), using simultaneous-equation models. Omri et al. [31] too analysed the causal interaction between $CO_2$ emissions and economic growth by using dynamic simultaneous equation data models and shown their positive impact on each other.

Chen et al. empirically demonstrated that per capita GDP is proportional to $CO_2$ emissions [53]. In the context of IRAN, Mousavi et al. [54] proved, using historical data, that increase in $CO_2$ emissions is largely because of economic activity. Chaabouni and Saidi [55] have experimentally proved a bidirectional causality between $CO_2$ emissions and GDP per capita. They did a case study of 51 countries and found that 1% increase in $CO_2$ emissions increases the economy by 0.011%. In addition, a growth of 1% in economy indicates an increase of 0.263% $CO_2$ emissions. Sugiawan et al. [74] analyzed data belonging to 105 countries to prove that any cut in $CO_2$ emission have a negative impact on the sustainable well-being of a nation. Can et al [75] highlighted that forecasted economic growth of India will quadruple $CO_2$ emission by 2050, indicating a strong positive effect of GDP on $CO_2$ emission. Acheampong [56] analysed the dynamic causal relationship among carbon emissions, energy consumption and economic growth in the context of 116 countries. The analysis results have depicted that carbon emissions positively cause economic growth. Stern [57] estimated that the carbon-income elasticity is 1.509 globally, suggesting that increase in carbon emission increases GDP. Wagner [58], [59] and Liddle [60] also pointed out that the per capita $CO_2$ elasticity of GDP is positive and monotonic. Sarkodie and Owusu [62] reported that, for Ghana, if energy-use, GDP, and population increases by 1%, then $CO_2$ emissions also increase by 0.58%, 0.73%, and 1.30% respectively. Marjanovic et al. [26] reported a novel way of predicting the GDP using $CO_2$ emissions from solid, liquid and gaseous fuel by using ELM. The present paper reports a DATL based GDP (termed DATL-GDP) prediction mechanism using $CO_2$ emissions of developed and developing economies.

## 3. Preliminaries

This section explains the preliminary concepts required to understand the proposed approach. Sub-section 3.1 gives an overview of transfer learning, and the remaining sub-sections provide insight into GRNN, ELM, and SVR respectively.

### 3.1 Transfer Learning

Transfer learning (TL) has been successfully utilized in various fields including failure prediction [5], and intelligent environments [38] etc. TL is also known by many other nomenclatures [6], [32] such as: transfer of learning, domain adaptation, multi-task learning, meta learning, etc. Various Transfer Learning approaches are depicted in Fig. 3 and are elaborately discussed in [25]. Some, related to this paper are explained below:

- **Domain:** A domain is represented by $\{F, P(X)\}$, where F denotes feature space and $P(X)$, $X = \{x_1, \ldots, x_n\}$, depicts every feature's marginal probability distribution. A feature's marginal probability is obtained by keeping all the remaining features constant.

- **Task:** A Task is defined by $T = \{Y, f(.)\}$ where $Y = \{y_1, \ldots, y_m\}$ depicts a label space. Predictive function is denoted by $f(.)$, which is trained using pairs $(x_i, y_i)$, and the labels for new instances are predicted by this learned function $f(.)$.

    **Transfer learning (TL):** For a source domain $D_s$, and its learning task $T_s$, a target domain $D_t$, and its learning task $T_t$, the rationale of TL is to model a learning function $f_t(.)$ in $D_t$, by utilizing $D_s$ and $T_s$, when $D_s \neq D_t$ or $T_s \neq T_t$.

Source domain refers to the data used for training and target domain refers to the testing data on which prediction is done. However, if both domains and their learning tasks are exactly similar i.e. $D_s = D_t$ and $T_s = T_t$, then the TL problem transforms into a machine learning problem.

- **Transductive TL:** In this case $D_s \neq D_t$. which implies either $F_s \neq F_t$ or $P_s(X) \neq P_t(X)$.
- **Domain adaptation technique:** This is a special case of transductive transfer learning, where the number of features are similar in source and target domain ($F_t = F_s$) but their marginal distribution differs i.e. $P_t(X) \neq P_s(X)$.
- **Sample selection bias:** A Transductive TL in which the generalization of results derived from small samples in order to predict the behaviour of the whole population, depends on the collected sample biasness [44].

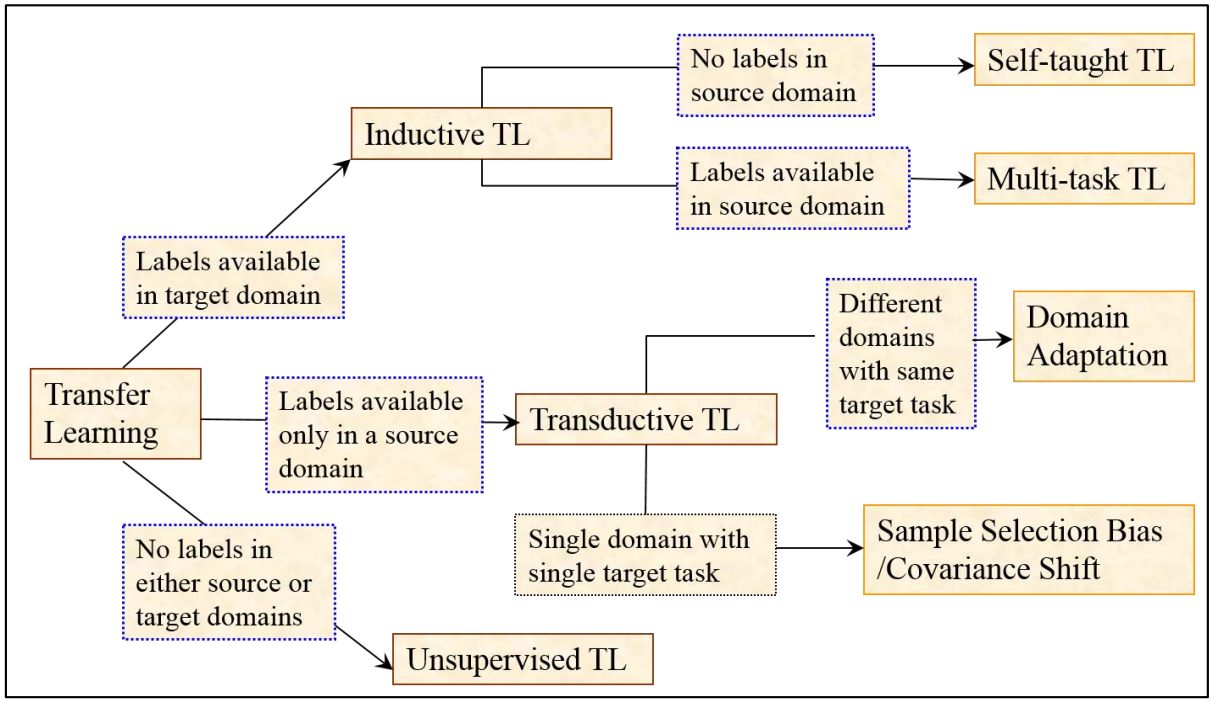

Fig. 3: Diverse Transfer Learning Approaches [32]

Various types of TL have evolved, depending upon different learning settings (see Table 2) and also the availability of class labels and type of tasks to be performed as depicted in the Table 3.

Table 2: TL specialization due to domains and task diversity [32]

| | Learning Settings | Source and Target Domains | Source and Target Tasks |
|---|---|---|---|
| | Traditional Machine learning | Same | Same |
| Transfer Learning | Inductive Transfer Learning | Same | Different but related |
| | Unsupervised Transfer Learning | Different but related | Different but related |
| | Transductive Transfer Learning | Different but related | Same |

Table 3: Labels availability in different TL approaches [32]

| TL Settings | Related Areas | Source Domain Labels | Target Domain Labels | Tasks |
|---|---|---|---|---|
| Inductive Transfer Learning | Multi-task Learning | Available | Available | Regression, Classification |
| | Self-taught Learning | Unavailable | Available | Regression, Classification |
| Transductive Transfer Learning | Domain Adaptation, Sample selection Bias, Co-variate shift | Available | Unavailable | Regression, Classification |
| Unsupervised Transfer Learning | | Unavailable | Unavailable | Clustering, Dimensionality Reduction |

### 3.2 Generalized Regression Neural Network

The GRNN proposed by Specht [40], is a special type of artificial neural network, which is not an iterative training method. It consists of four layers: (i) input layer, (ii) pattern layer (first hidden layer), (iii) summation layer (second hidden layer), and (iv) output layer, as shown in Fig. 4. Input layer has nodes equal to the number of features in the data set. The number of nodes in the pattern layer are equal to the number of data points in the training data set. Every node of the pattern layer represents a non-linear transformation with a distinct training data point. The summation layer has two nodes, which represent the numerator and the denominator of the Eq. (1) respectively. The output layer produces the result, denoted $\bar{Y}$.

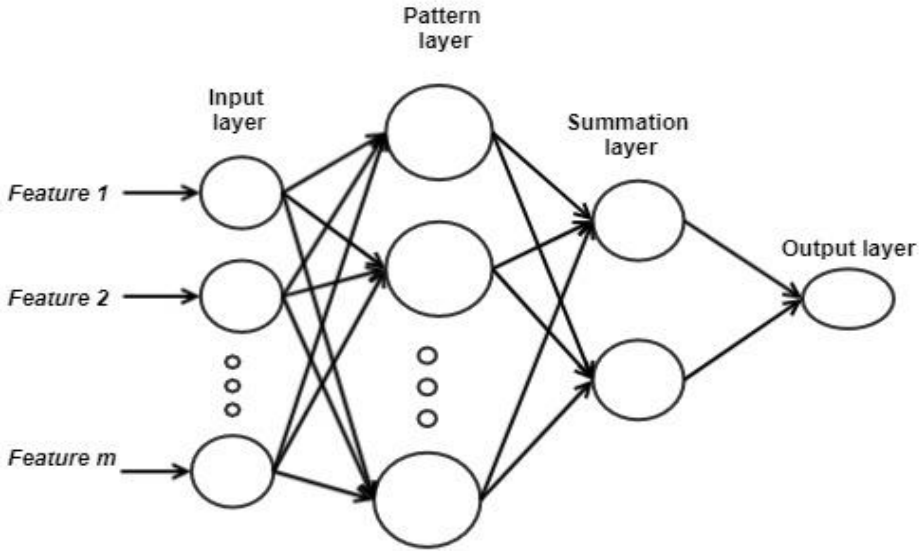

Fig. 4 GRNN Architecture

For an input vector X, GRNN produces an output $\bar{Y}$ as follows:

$$\bar{Y}(X) = \frac{\sum_{i=1}^{n} Y_i * exp(-S_i/2\sigma^2)}{\sum_{i=1}^{n} exp(-S_i/2\sigma^2)} \tag{1}$$

Here, $S_i = (x - x_i)^T(x - x_i)$, $n$ denotes the total number of observations, and $\sigma$ is the spread parameter. The pair of $(x_i, y_i)$ denotes the $i^{th}$ training data point and its label, respectively. To fit the data precisely, the spread parameter $\sigma$ should be chosen as slightly smaller than the average of the distance between the provided data points. If $\sigma$ is too large, then $\bar{Y}(X)$ will be the mean of all the training labels, and if $\sigma$ is too small, then $\bar{Y}(X)$ will assume label value of closest training data points.

GRNN does the approximation of any random function between the input and output vectors by making an estimate from the training data directly. GRNN is highly consistent as the training set size is inversely proportional to the estimation error, i.e., estimation error tends to zero as training set size grows. As with regression methods, GRNN too, estimates continuous variables. It is allied to the radial basis function network and uses the kernel regression technique, a standard statistical technique. GRNN derives the pdf (joint probability density function), using only the training set [23]. The benefits of GRNN are as follows:

a) The system is perfectly general as pdf is estimated from the data itself.
b) Regression surfaces are constructed without deriving any assumptions about the nature of the prediction model.

*3.3 Extreme Learning Machine*

Extreme Learning Machine (ELM), a learning algorithm based on single hidden layer feed forward neural networks (SLFNs) was proposed by Huang et al. [1], [21]. A typical ELM architecture is shown in Fig. 5. It learns the model swiftly compared to neural networks and support vector machines. ELM generalization performance has proved to be better than gradient-based learning in most experiments. It is capable of addressing many demerits of gradient-based learning algorithms, e.g., trapping in local minima, improper learning rates, and over-fitting. For many non-differentiable activation functions, ELM learning algorithm can be exploited to train SFLNs, while gradient-based learning algorithms work for differentiable activation functions only [22]. ELM input weights and hidden layer biases are randomly assigned to hidden layer nodes. The output layer weights are analytically deduced using least square method and Moore–Penrose generalized inverse of the hidden layer output matrix.

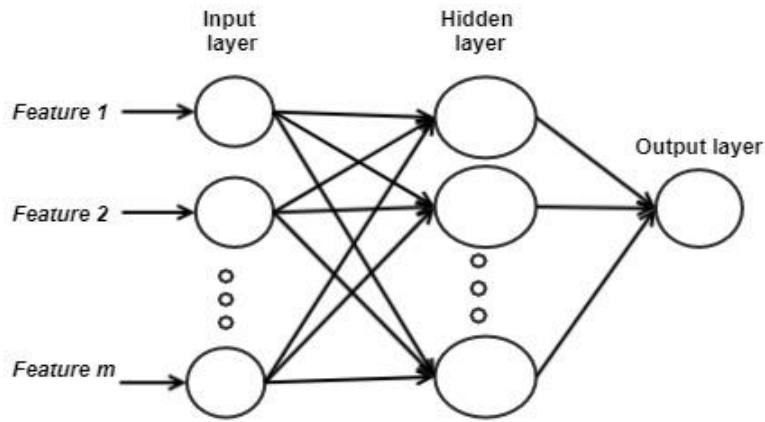

Fig. 5 ELM Architecture

In ELM, input data is first mapped to $M$ dimensional random space. The ELM output $f_M(x)$ is calculated as follows:

$$f_M(x) = \sum_{k=1}^{M} \omega_k h_k(x) = \omega * h(x) \qquad (2)$$

Here, $h(x) = [g_1(x), \ldots, g_M(x)]$ are the output of hidden nodes with random weights; $g_k(x)$ is the output of the k[th] hidden node for an input $x$, $g()$ being the activation function.

The output weight matrix between the hidden and the output nodes is represented by $\omega = (\omega_1, \ldots, \omega_M)^T$. For $N$ training data $(x_l, y_l)_{l=1}^{N}$, $x_l$ and $y_l$ respectively denote the l[th] training data point and its target label, ELM solves the following problem:

$$\omega * H = Y \qquad (3)$$

Here, $H = (h(x_1), \ldots, h(x_N))^T$ is the hidden node output matrix and $Y = [y_1, \ldots, y_N]$ is the label. The output weights are analytically computed using Eq. (4)

$$\omega = Y * H\hat{} \qquad (4)$$

where, $H\hat{}$ is the Moore-Penrose generalized inverse of $H$ matrix.

In this paper, we have used the Kernel-ELM method [49]. Kernel-ELM differs from ELM in the way hidden layer is computed. In hidden layers of Kernel-ELM, random weights are not generated, rather kernel weights are computed using Eq. (5) with respect to each training data acting as a node in the hidden layer.

$$h(x_j) = K(x_1, x_j), K(x_2, x_j), \ldots, K(x_N, x_j) \qquad (5)$$

The computation of the final layer is similar to that of ELM using Eq. (2-4).

*3.4 Support vector machine (SVM) and support vector regression (SVR)*

SVM, proposed by Vapnik et al. [43], is a famous "maximum margin" classifier. It maximizes the margin between two separating hyperplanes. By maximizing the margin, it attempts to minimize the generalization error [10], [12]. It is an optimization task, which minimizes the convex quadratic function subject to linear inequality constraints.

Support Vector Regression (SVR) is a method to solve the regression problem. It is analogous to the SVM in principle. It is also an optimization task in which a convex quadratic function is minimized subject to linear inequality constraints [8]. The dual of this minimization problem is given by the following Eq. (6).

$$\text{Maximize}_{\gamma,\theta} \quad Y^T(\gamma - \theta) - (\gamma - \theta)^T K (\gamma - \theta) - \epsilon \ddot{\omega}^T(\gamma + \theta) \qquad (6)$$

$$\text{Subject to} \quad \ddot{\omega}^T(\gamma - \theta) = 0,$$

$$0 \leq \gamma, \theta \leq C$$

Here, $K$ is the kernel matrix, whose elements are $K(x_i, x_j) = \emptyset(x_i) * \emptyset(x_j)$. Nonlinear transformation is denoted by $\emptyset : R^n \to R^m$, which transforms the data points from input space ($R^n$) to a higher dimension feature space ($R^m$). For $N$ training data $(x_l, y_l)_{l=1}^{N}$, $x_l \in R^n$ and $y_l \in R$ respectively denotes the l[th] training data point and its target label. So, $Y$ is a vector consisting of all the labels, i.e. $Y = \{y_1, y_2, \ldots, y_N\}$. Also, $\ddot{\omega}$ is a vector of 1's, which is used for summation. The constant $C$ is the regularization coefficient for controlling overfitting as it imposes a penalty on observations lying out of the epsilon (*ε*) margin. All errors lying within *ε* distance of the observed value are ignored by the linear ε-insensitive loss function, as it treats them equal to zero [50].

## 4. Proposed Methodology

In this section, we shall explain our proposed transfer learning approach to predict per capita GDP of a country using its $CO_2$ emission data. For better understanding of the proposed approach, we first briefly discuss about the dataset. Then, the proposed methodology is elaborately explained followed by a brief discussion of the performance evaluation criteria.

*4.1 Dataset details*

The dataset for this experiment has been taken from the World Bank database for the USA, the European Union, India and Cameroon [51]. These datasets comprise values collected over a span of 54 years, i.e., from 1960 to 2013. This data was accessed in November 2016. $CO_2$ data has been analysed for emissions from gaseous, liquid, and solid fuel consumptions separately, as well as the total $CO_2$ emissions in metric tons per capita.

Table 4: Input parameters

| Sr. No. | Parameter's Description |
|---------|-------------------------|
| Input 1 | $CO_2$ emissions from gaseous fuel consumption (% of total) |
| Input 2 | $CO_2$ emissions from liquid fuel consumption (% of total) |
| Input 3 | $CO_2$ emissions from solid fuel consumption (% of total) |
| Input 4 | $CO_2$ emissions (metric tons per capita) |

$CO_2$ emissions in metric tons per capita have been used for efficient transfer learning, as percentage does not define a unique value during knowledge transferral in DATL-GDP. GDP per capita at current US dollars (US$), is the output since per capita GDP also incorporates the population size of that country.

Table 5: Output parameters

| Sr. No. | Parameter's Description |
|---------|-------------------------|
| Output 1 | GDP per capita (at current US$) |

The input and output parameters used in this research are explained in Table 4 and Table 5 respectively. For the better analysis of the outcome of our proposed approach, we have considered data belonging to a diverse set of countries [66] as follows:

(i) Developed economies viz., United States of America (USA) and the European Union (EU)

(ii) Developing economies viz., India (IND)

(iii) Highly indebted poor or least developed economies viz., Cameroon (CMR)

The logic behind the choice of the above countries is that they all are connected to the sea (i.e., are not landlocked), have wide differences in their $CO_2$ emissions and their GDPs. They also have a population of at least 20 million, and are situated in different continents. Hence, they are gigantically diverse from each other and thus best suited for our proposed DATL.GDP experimentations.

*4.2 Model training scheme via transfer learning*

This section describes the working of our proposed GDP prediction methodology, which we have termed as "Domain Adapted TL for GDP prediction (DATL-GDP)". It is diagrammatically depicted in Fig. 6. Algorithm 1 shows the steps involved in the proposed DATL-GDP.

---

**Algorithm 1**: DATL-GDP ($D_m, T_m, D_t, T_t$)

**Inputs**: $D_m$ – Mixture of source domain data and little amount of target domain data,

$T_m$ – Per capita GDP values corresponding to $D_m$, $D_t$ − Target domain data,

$T_t$ − Per capita GDP values of $D_t$.

**Output**: $T'_t$ − Predicted per capita GDP values.

**Begin**
{
    Step 1: Construct $D_m$

    Step 2: **Do** Regression

    Step 3: Train regression model on $D_m$ using labels $T_m$, to predict

        per capita GDP

    Step 4: Use trained regression model to output $T'_t$ for $D_t$

    Step 5: Performance evaluation of $T'_t$ with respect to $T_t$

    Step 6: **Return** $T'_t$
}
**End**

---

In the Step-1, DATL-GDP constructs training data ($D_m$), which is a mixture of complete source domain data and little amount of target domain data. $D_m$ is of four dimension, having four parameters as described in Table 4. To train a regression model, output label (per capita GDP) values are also needed with respect to $D_m$; these output labels are denoted by $T_m$ (as described in table 5). In Step-2 choose any of the regression techniques from GRNN, ELM, or SVR, which will then be used in later steps. In Step-3, using $D_m$ as input vector and $T_m$ as labels, train the chosen regressor. This trained regressor is later tested using target domain data ($D_t$) in Step-4. Parameter dimensions of $D_t$ is similar to that of $D_m$ as explained in Table 4. Per capita GDP values predicted by the regressor over $D_t$ are denoted by $T'_t$. Finally, the prediction is evaluated by employing performance criteria (see Subsection 4.3) over the predicted $T'_t$ and $T_t$, where $T_t$ is the actual per capita GDP values corresponding to $D_t$) and both $T'_t$ and $T_t$ have parameters similar to that of the $T_m$ (Table 5).

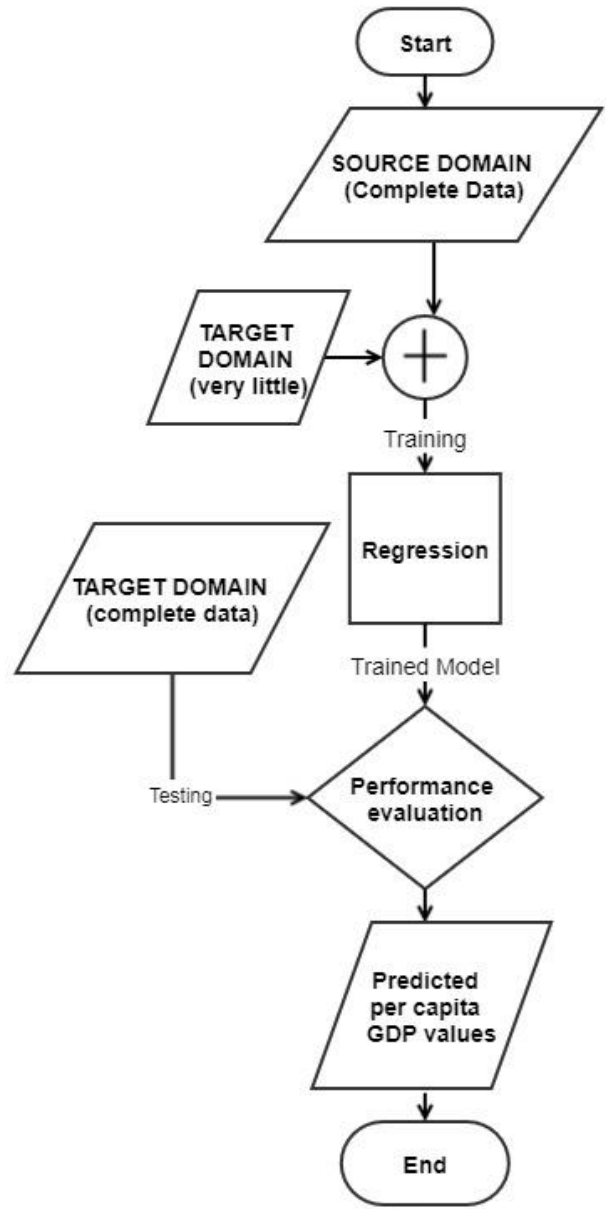

Fig. 6: Flowchart of the proposed GDP prediction approach

The proposed model is called 'Domain adapted TL for GDP prediction', because the distribution of the source domain data (training data) and the target domain data (testing data) is gigantically dissimilar to each other. For example, let the source domain is the $CO_2$ emission data of USA and the target domain is the $CO_2$ emission data of INDIA. It is well-known that both of these countries are highly diverse in terms of development, geographical location, education, etc. Hence, marginal probability distribution of source domain ($D_s$) data and target domain ($D_t$) data is extremely different, i.e., $D_s \neq D_t$ due to the $P_t(X) \neq P_s(X)$, which is a perfect case of domain adaptation in transductive TL. In this scenario, as the model is trained using the $CO_2$ emission dataset of a developed country (USA) to predict its GDP; then the model is specifically adapted to predict GDP of USA or any other country whose $CO_2$ emission dataset and GDP is similar to that of the USA (i.e. training data). However, if the model is tested on the $CO_2$ emission data of a developing country (INDIA) whose dataset distribution is highly dissimilar to that of USA (training data), then the model's prediction accuracy will be very poor. It happens because the testing data came from an unseen distribution about which the trained model has no prior knowledge.

This proposed technique is employed to estimate the per capita GDP of those countries whose values are missing in the dataset of the World Bank [51]. It is not feasible to train a model for each of these countries using only their small amount of available data since in order to train a ML model, a large amount of data is required. Therefore, the limitations of a model's performance on a highly dissimilar dataset than the dataset with which it is trained and non-availability of sufficient data for training, motivated us to propose the DATL-GDP approach. It aims to tackle distribution differences among the source and target domains, by designing a robust and adaptive 'predictive model' imbibing the characteristics of TL. In addition, the DATL-GDP approach can be effectively put to use in conjunction with the existing methodologies, for estimating the per capita GDP of a country by using only its $CO_2$ emissions data.

*4.3 Performance Evaluation*

The predictions of each model are evaluated for its preciseness using root mean square error (RMSE) and Coefficient of determination ($R^2$). Both of the statistical methods are defined below:

*A) Root Mean Square Error (RMSE):* Let $\bar{Y}_i$ be the predicted value, $Y_i$ be actual value and $N$ be the number of observations. Then RMSE can be calculated using Eq. (7):

$$\text{RMSE} = \sqrt{\left(\frac{\sum_{i=1}^{n}(\bar{Y}_i - Y_i)^2}{N}\right)} \tag{7}$$

The model prediction accuracy is inversely proportional to the RMSE value, i.e., the designed model is very good if RMSE is minimum. Relative Root Mean Square Error (RRMSE) [29] is RMSE divided by the mean of the actual data.

*B) Coefficient of Determination ($R^2$):* $R^2$ statistically measures the perfection of the regression line in approximating the actual data points. It may be computed using the mathematical formula given in Eq. (8).

$$R^2 = 1 - \frac{\sum_{i=1}^{n}(\bar{Y}_i - Y_i)^2}{\sum_{i=1}^{n}\left(Y_i - \frac{\sum_{j=1}^{n} Y_j}{N}\right)^2} \tag{8}$$

A predictive model is considered to be excellent whenever the coefficient of determination ($R^2$) approaches 1. It performs poorly if $R^2$ decreases from 1.

## 5. Experimental results and performance analysis

This proposed framework has been implemented using MATLAB on a Xeon® processor with 48 GB RAM on Windows 7 OS. For the experimentation, we randomly chose a country as source domain and the rest of the countries as target domains. For example, we chose

CAMEROON as the source domain and EUROPE as the target domain and termed this experiment "CMR-to-EU". The prediction of GDP by GRNN, SVR and ELM is shown in Fig. 7 (a) with the title: "CMR-to-EU ". Like this, a total of 12 experiments were performed as follows: (i) CMR-to-USA, (ii) CMR-to-IND, (iii) CMR-to-EU, (iv) IND-to-USA, (v) IND-to-EU, (vi) IND-to-CMR, (vii) EU-to-CMR, (vii) EU-to-IND, (ix) EU-to-USA, (x) USA-to-EU, (xi) USA-to-CMR, (xii) USA-to-IND. These 12 experiments represent a different pairwise combination (i.e. source and target data) of DATL-GDP among CMR, EU, IND, and USA. Each of these 12 experiments may be placed under two different categories. In the first category, source domain is a developing country while in the second category, source domain is a developed country. Results from the two categories are shown in Fig. 7 and Fig. 8, respectively.

For each of these 12 experiments, the prediction accuracy of all the three classifiers (GRNN, SVR and ELM) has been analysed on the basis of their root mean square error (RMSE) shown in Fig. 9 and coefficient of determination ($R^2$) shown in Fig. 10. The prediction is termed very accurate if the RMSE value of an experiment is small and its $R^2$ value is near to 1. In nearly all cases, SVR performed the worst among all, with exorbitantly high RMSE and very small $R^2$ values. ELM performed better than SVR in almost all experiments but one, ("IND-to-CMR") by predicting smaller RMSE values. For 10 experiments out of 12, GRNN had the highest $R^2$ values as compared to SVR and ELM. The remaining 2 experiments were very close to the maximum values. However, the analysis of RMSE plot in Fig. 9 proves that GDP predicted by GRNN for these countries is the most accurate as compared to those by SVR and ELM.

In each of these 12 experiments, the difference between the GDP of the source domain and the GDP of the target domain was calculated for all the values of their GDPs from the year 1960 to 2013 (54 years in total). Then the average of these differences was plotted in ascending order as shown by the 'red line' in Fig. 11, where the corresponding values of the 'blue line' depict the GDP prediction error of the GRNN model with respect to the 'red line'.

Also, we can deduce some interesting observations from the Fig. 11, as summarized below:

a) Whenever a rich country like the USA or the EU is chosen as the source domain, and the target domain is a developing (IND) or a least developed country (CMR), then the prediction error is minimum.

b) Smaller the GDP difference among the countries, less is the GDP prediction error. As an experiment, India's data was considered as the source domain and the target domain was the data of Cameroon. Since, their GDP difference is very small, the error (RMSE) value for GDP prediction also turns out to be minimal or vice versa.

Prediction accuracy is very poor, if a model is trained on a developing (say IND) (or least developed country (say CMR)) and tested to predict the GDP of a rich country (say USA). Therefore, we can infer that transfer learning is inefficient if source domain is a poor country and target domain is a rich country. It may be because per capita GDP of developed countries during initial years (1960 onwards) are somehow at par with the recent per capita GDP values of the developing countries. Whereas, developing countries have never witnessed in its entire past higher GDP values similar to those of developed countries. Due to this, TL is comparatively better when source domain is a developed country and target domain is a developing country; while TL performance degrades when it is vice versa.

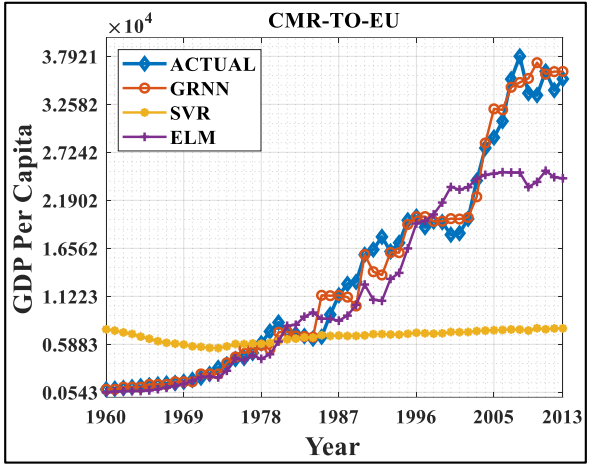
(a)

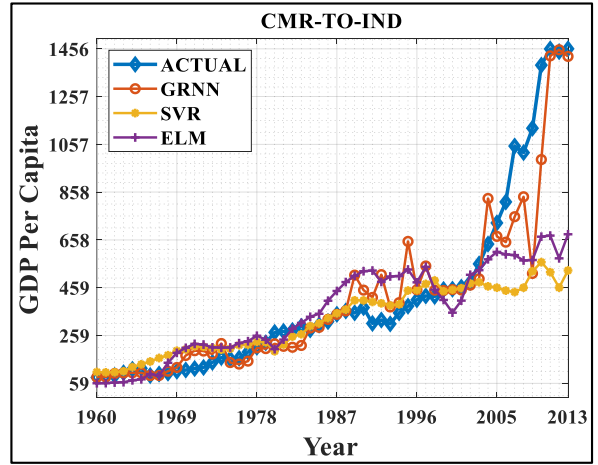
(b)

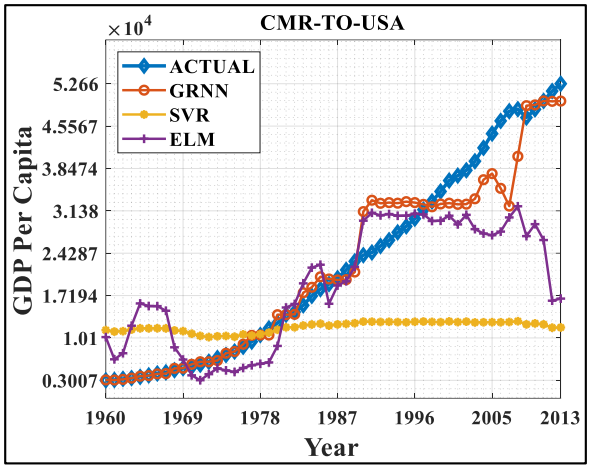
(c)

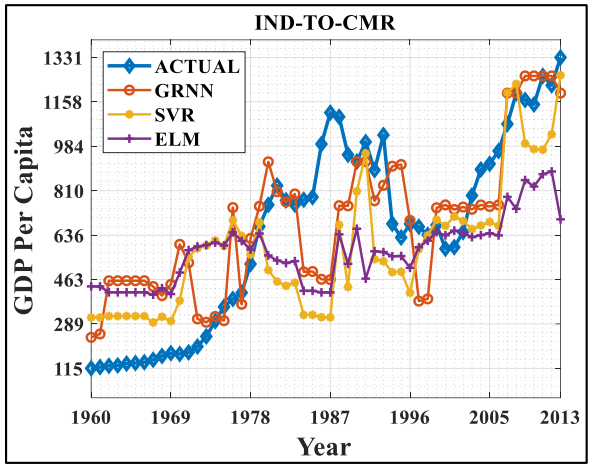
(d)

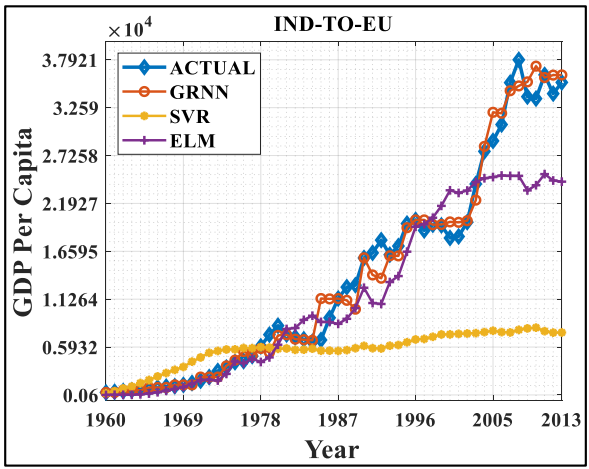
(e)

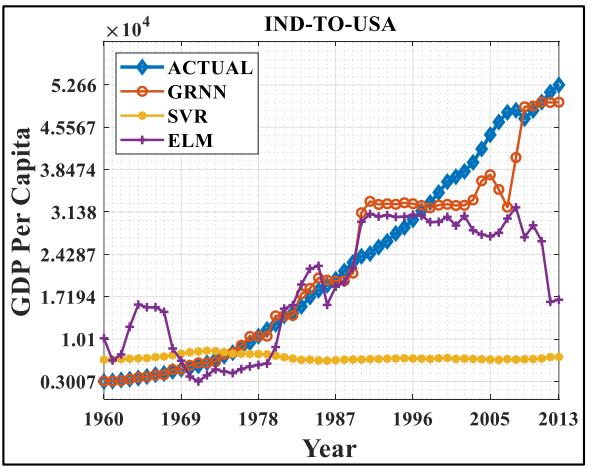
(f)

Fig. 7: Results of the proposed DATL-GDP experiments when the source domain is a developing country and target domain is a developing/developed countries: (a) Source domain CMR, Target domain EU, (b) Source domain CMR, Target domain IND, (c) Source domain CMR, Target domain USA, (d) Source domain IND, Target domain CMR, (e) Source domain IND, Target domain EU, (f) Source domain IND, Target domain USA

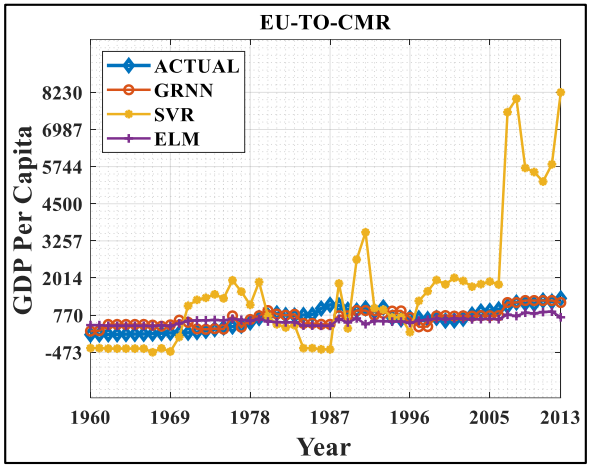
(a)

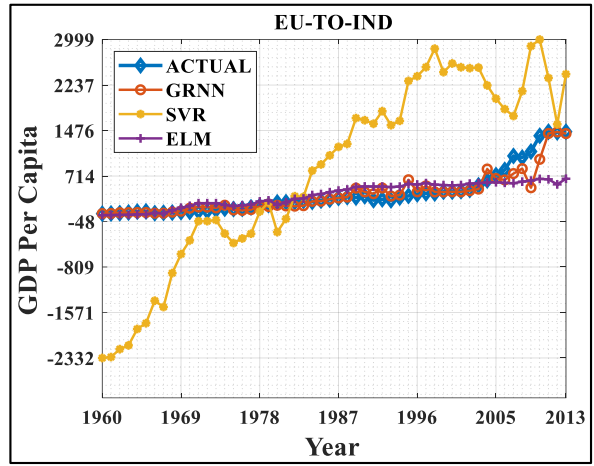
(b)

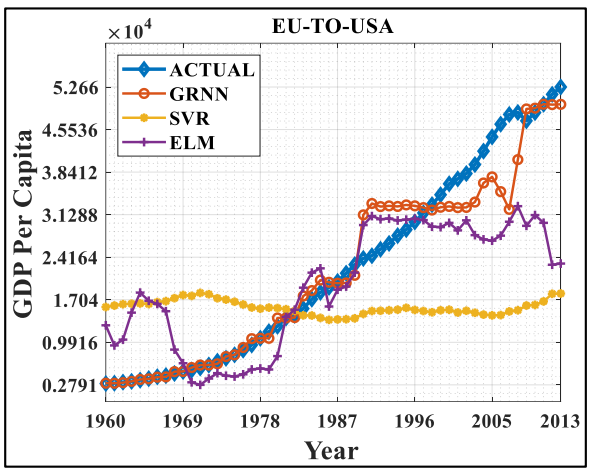
(c)

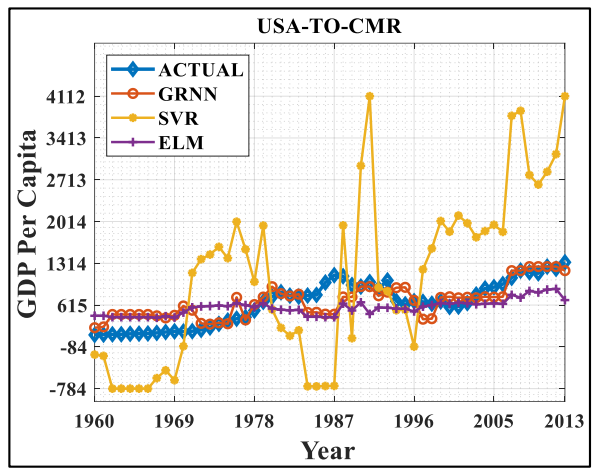
(d)

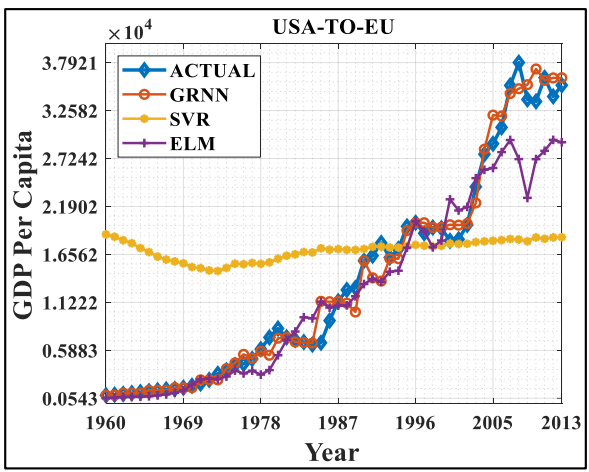
(e)

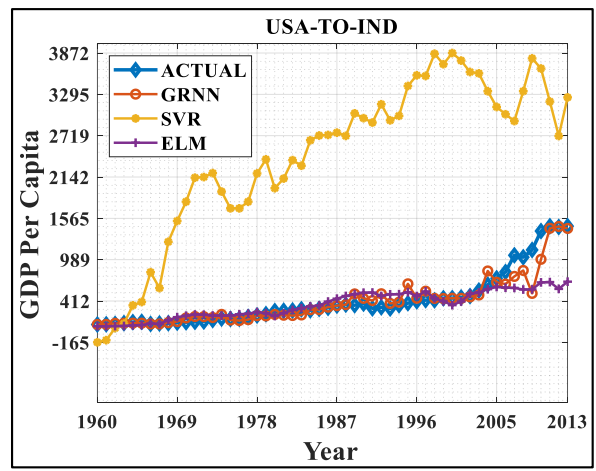
(f)

Fig. 8: Results of the proposed DATL-GDP experiments when the source domain is a developed country and target domain is a developing/developed countries: (a) Source domain EU, Target domain CMR (b) Source domain EU, Target domain IND, (c) Source domain EU, Target domain USA (d) Source domain USA, Target domain CMR, (e) Source domain USA, Target domain EU (f) Source domain USA, Target domain IND

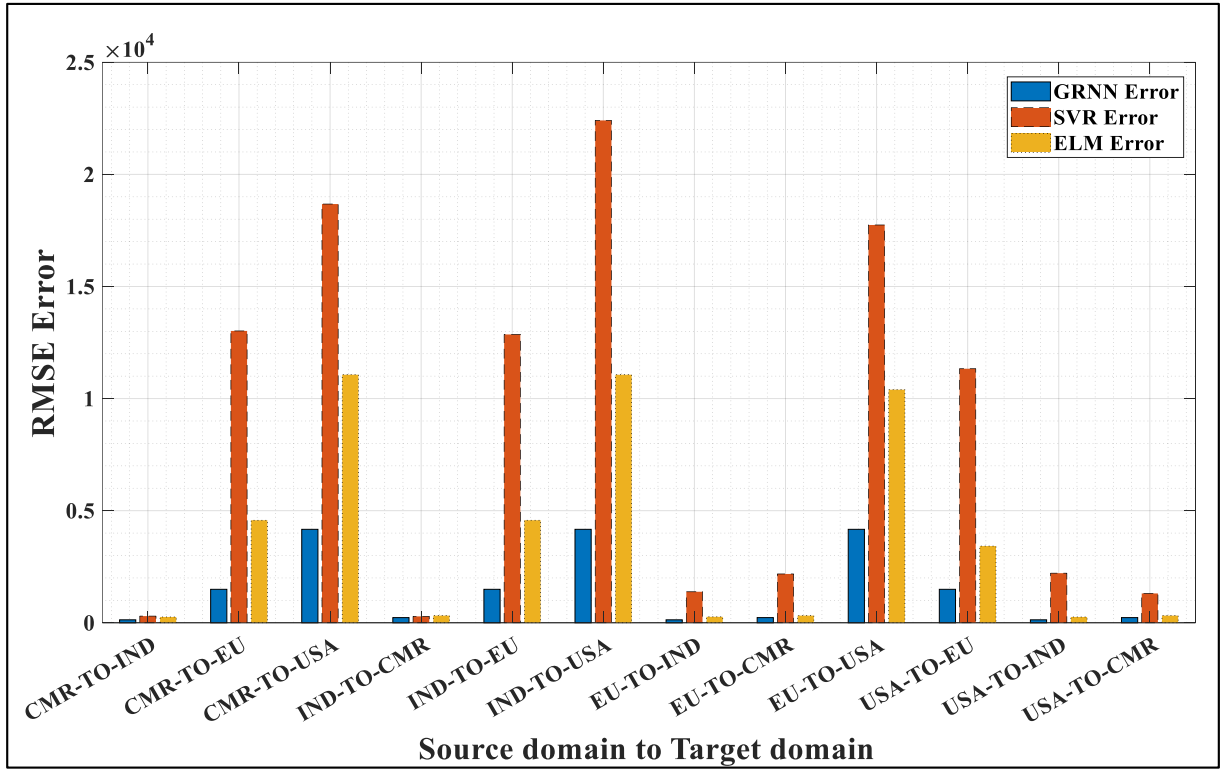

Fig. 9: RMSE values for all the 12 experiments

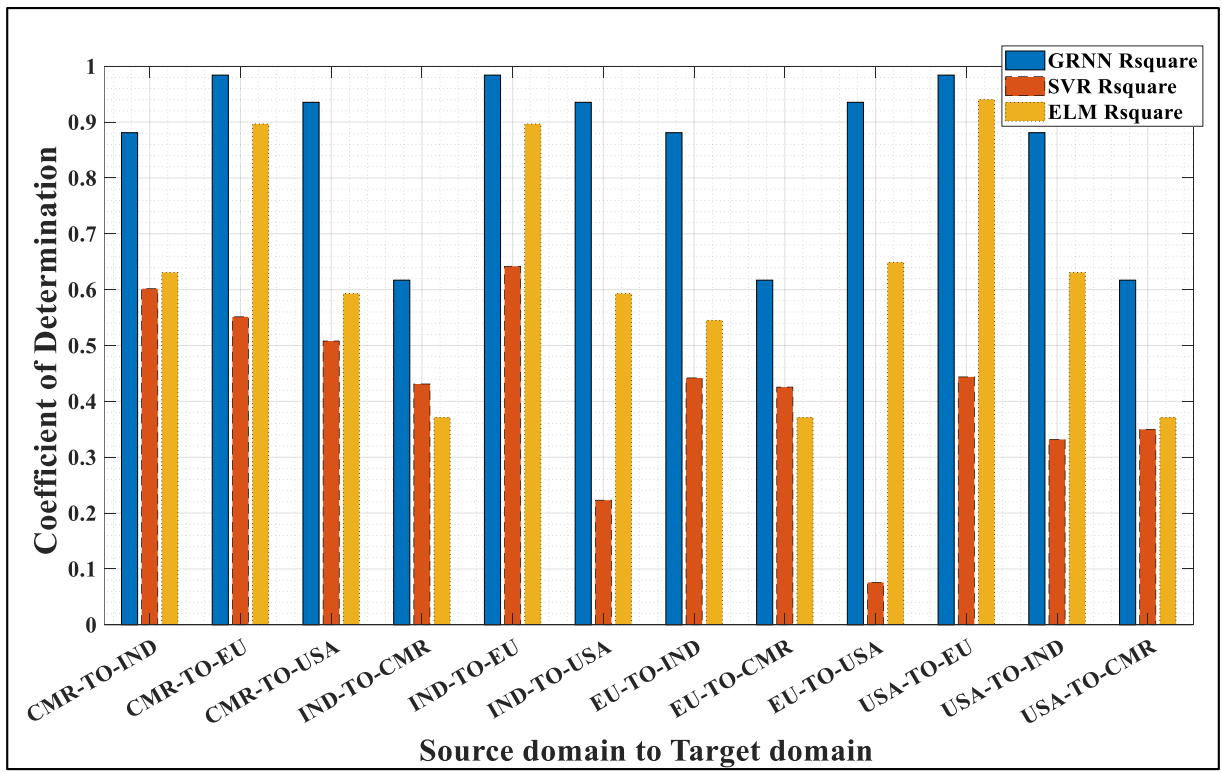

Fig. 10: Coefficient of determination values for all the 12 experiments
18

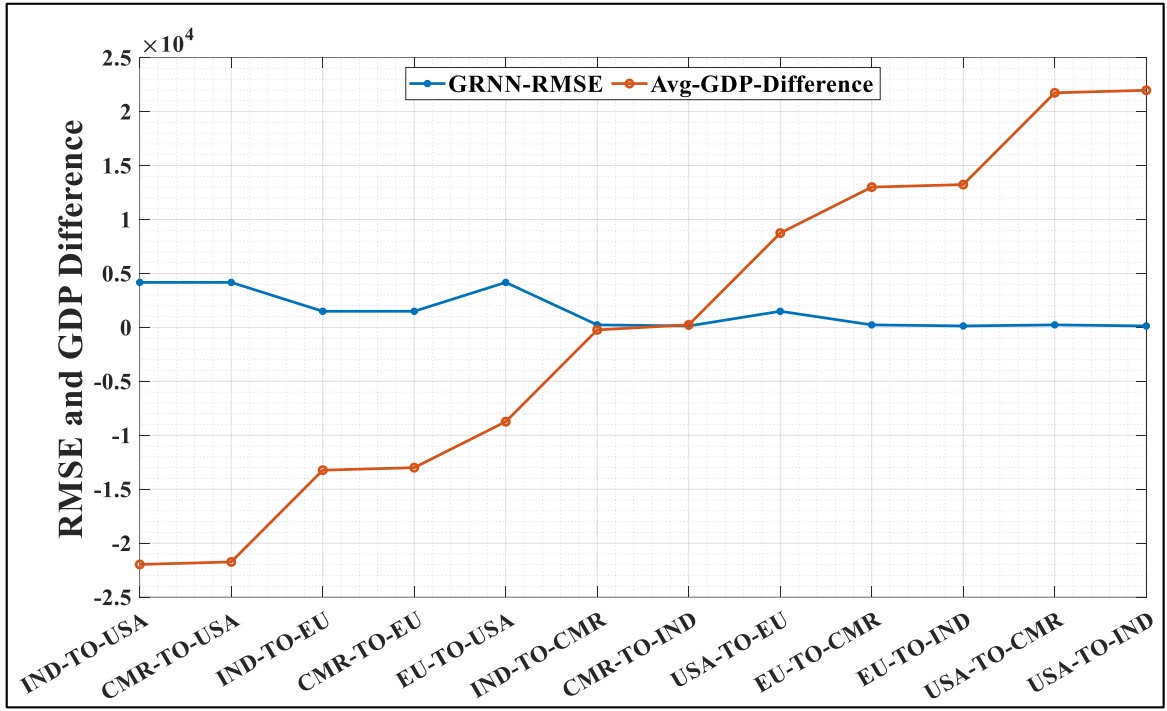

Fig. 11: RMSE versus Average GDP difference of each of the 12 experiments

c) Exception: An anomaly is observed when the GDP of the EU is predicted by a model trained on a developing or developed country in the experiment. Here, the prediction error is comparatively small or nearly equal, i.e., RMSE obtained in the experiments of the IND-to-EU or CMR-to-EU are almost comparable. The possible reason behind this exception could be that the EU comprises 28 highly diverse nations, with countries like Luxembourg having nearly double the per capita GDP of the USA, while other European Union countries like Bulgaria and Romania have a GDP that is not even double that of India. This diversity among the countries of the EU possibly results in better transfer learning in predicting EU GDP when source domain is either INDIA or CAMEROON.

These observations vindicate our claim that the GDP of war-torn countries like the Syria, Afghanistan and Iraq, or for countries whose GDPs are unknown at any time period, can reliably be estimated by DATL-GDP, even when their available macroeconomic data is not sufficient for the precise calculation of their GDP.

## 6. Benefits of DATL-GDP

In this section, we try to highlight the benefits of the proposed DATL-GDP approach. In the previous section, 33.3% of the target data was used in DATL-GDP. The total amount of data available with regard to these countries was of 54 years, and 33.3% corresponds to one-third of the total data, i.e., of 18 ($= \frac{1}{3} \times 54$) years. This section experiments with different values of the target data to prove the significant improvements in DATL-GDP.

A number of experiments were conducted to elaborate the proposed methodology by imbibing different percentages of the target data during training.

For all of these twelve experiments, we did six different studies of TL, termed as: (i) *'No_TD'*, (ii) *'1/18*TD'*, (iii) *'1/9*TD'*, (iv) *'1/6*TD'*, (v) *'1/3*TD'* and (vi) *'1/2*TD'*. The average values of the RMSE obtained from these six studies are depicted in the Fig. 12. In the first study *'No_TD'*, GRNN, ELM and SVR were trained using only the source domain data, where no amount of target domain data was mixed. Then, the trained models (GRNN, ELM and SVR) were tested on the complete target domain and average RMSE values were calculated. In the second study *'1/18*TD'*, the training data contained a mixture of all the source domain data and one-eighteenth of the target domain data. That is, 3 years (54 years divided by 18) of target data points were mixed with the complete source domain data for training. Then, average RMSE values were calculated after testing the models on the complete target domain.

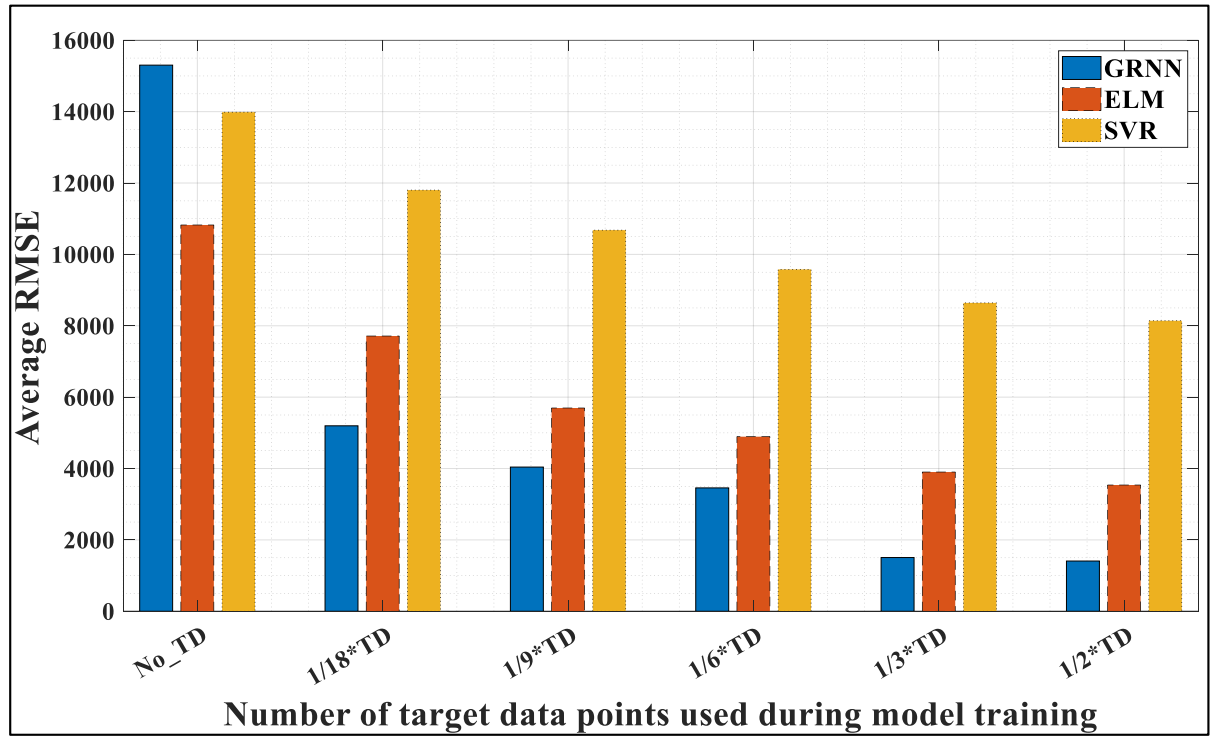

Fig. 12: Average RMSEs for different combinations of training data

Similarly, for the third study *'1/9*TD'* training was done on the collection of all the source domain data along with one-ninth of the target domain data, which means that 6 years (i.e. 54 years divided by 9) of the data points were taken from target domain together with the complete source domain data for training. Then, the testing of the trained models was conducted on the complete target domain and the average RMSE values were calculated. Like-wise, for the remaining three studies *'1/6*TD'*, *'1/3*TD'* and *'1/2*TD'*, one-sixth, one-third and one-half of the target domain data were respectively mixed with the source domain data for training. These studies were conducted upto *'1/2*TD'* (50 % of the target data mixing), so that the DATL-GDP model could be rigorously tested with a significant portion of unseen data.

Analysis of the results of these six studies are depicted in Fig. 12. As in '*No_TD'* case (the first study), ELM gives the best prediction accuracy with minimum error, while GRNN performed the worst. However, in the case of '*1/18*TD'* (second study), which considered the addition of just three years data points from the target domain (with the source domain data) while training, the GRNN, ELM and SVR improves its prediction accuracy by almost 66%, 28.7% and 15.6%, respectively. It can be verified that as the number of target data points during training keep on increasing in the Fig. 12 from left to right, the RMSE of GRNN, ELM and SVR goes on decreasing, i.e., the prediction accuracies of these models increases vastly from left to right. From *'No_TD'* study to *'1/2*TD'* study the improvement in prediction accuracy of GRNN, ELM and SVR is by at least of 90%, 67% and 41% respectively, as depicted in Fig. 13. Hence, it can be seen that as number of target domain data increases in DATL-GDP, the adaptive capability of regression techniques also increases enormously. It can be verified from Fig. 12 that from *'1/3*TD'* to *'1/2*TD'*, reduction in the RMSE is negligible, which symbolizes that the prediction performance converges around *1/3*TD*. Therefore, we have chosen 33.3% of the target data (i.e. *'1/3*TD'*) for training to depict the detailed experiments of the proposed DATL-GDP approach.

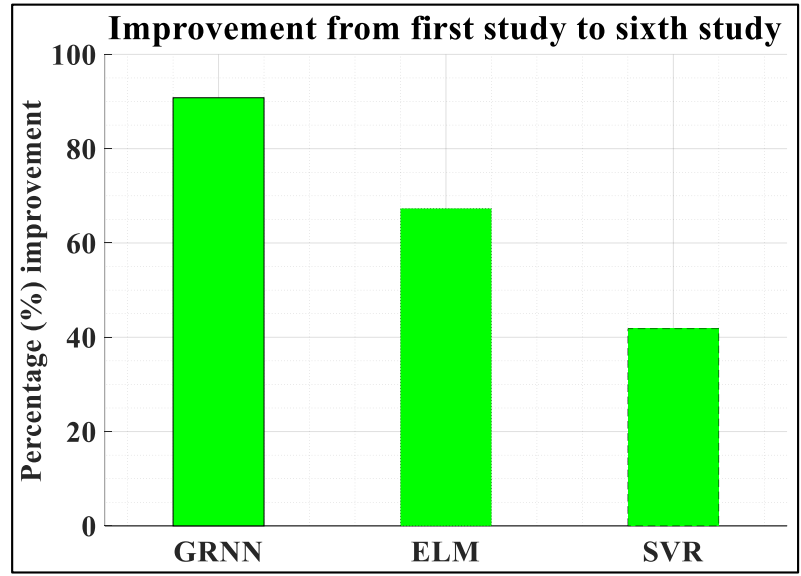

Fig. 13: Prediction improvement from *No_TD* to *1/2*TD*

### 6.1 Application of DATL-GDP approach

In this section, DATL-GDP has been used to predict the missing GDP of many countries such as Afghanistan, Iraq, Poland, Myanmar, Syria, Switzerland and Yemen [51]. These countries are highly dissimilar from each other as Poland and Switzerland are developed economies, Iraq, Syria, Myanmar, Yemen and Afghanistan are developing economies [66]. Among these, the last three countries are least developed and last one (Afghanistan) is highly in debt. In addition, Iraq, Syria, Yemen and Afghanistan are the world's most war-torn countries and Myanmar is largely isolated, due to which their macroeconomic statistics are not reliably available for the accurate calculation of their GDP. While conducting the experiments, we have clubbed these countries in two groups: isolated/war-torn countries and developed countries with missing GDP data. Results from the two groups are represented in Fig. 14 and Fig. 15, respectively.

It can be verified from the link in [51] that the GDP of these countries have a lot of missing values at various stages since 1960 onwards, as displayed in the left side of the Fig 14 and Fig. 15. The missing time periods of these countries are also given in the Table 6. To predict these GDP accurately, we used the DATL-GDP model shown in Fig. 6. To perform these experiments reliably, one third of the available GDP data of these countries was used as validation dataset.

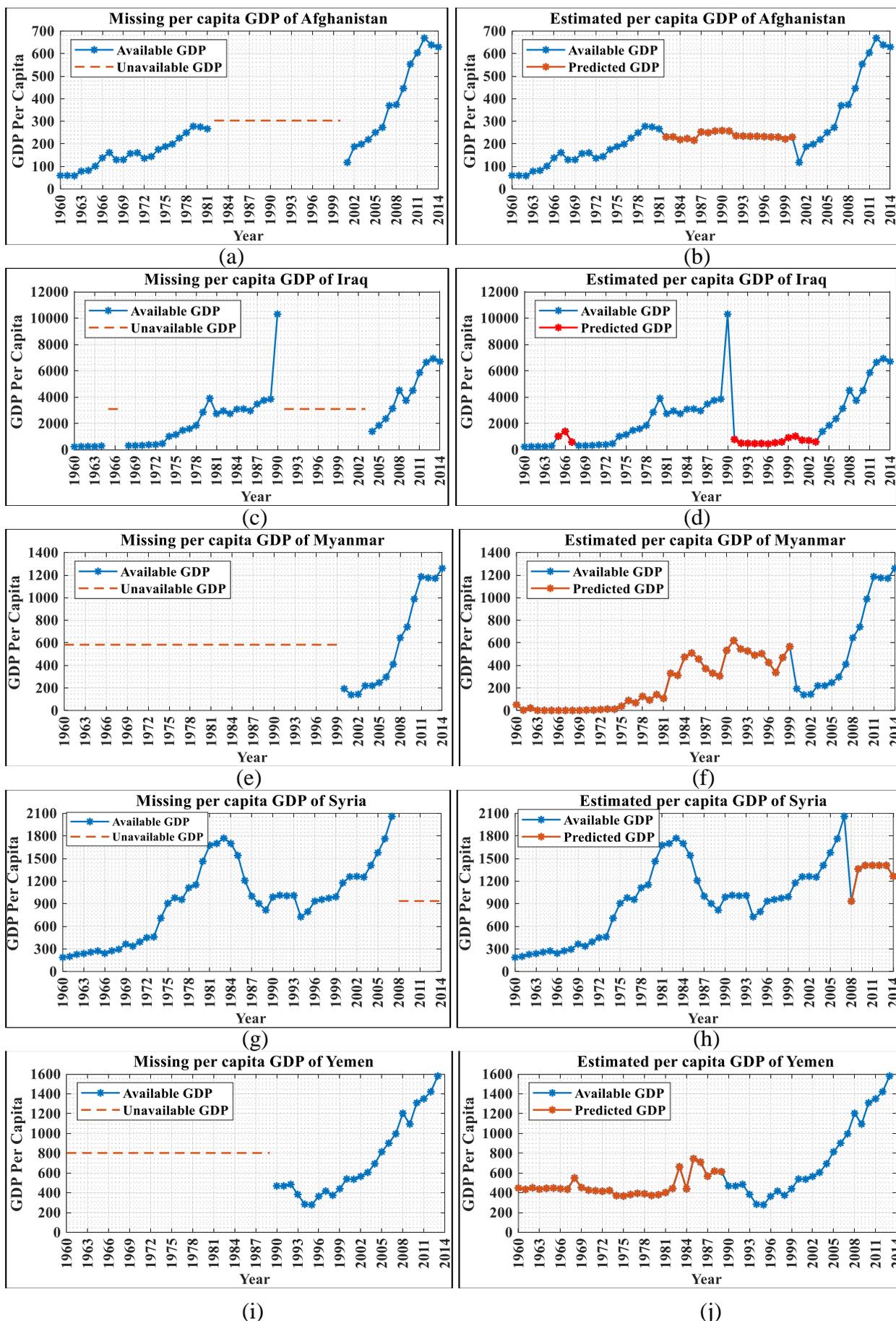

Fig. 14 Isolated/war-torn developing countries with their missing and estimated per capita GDP

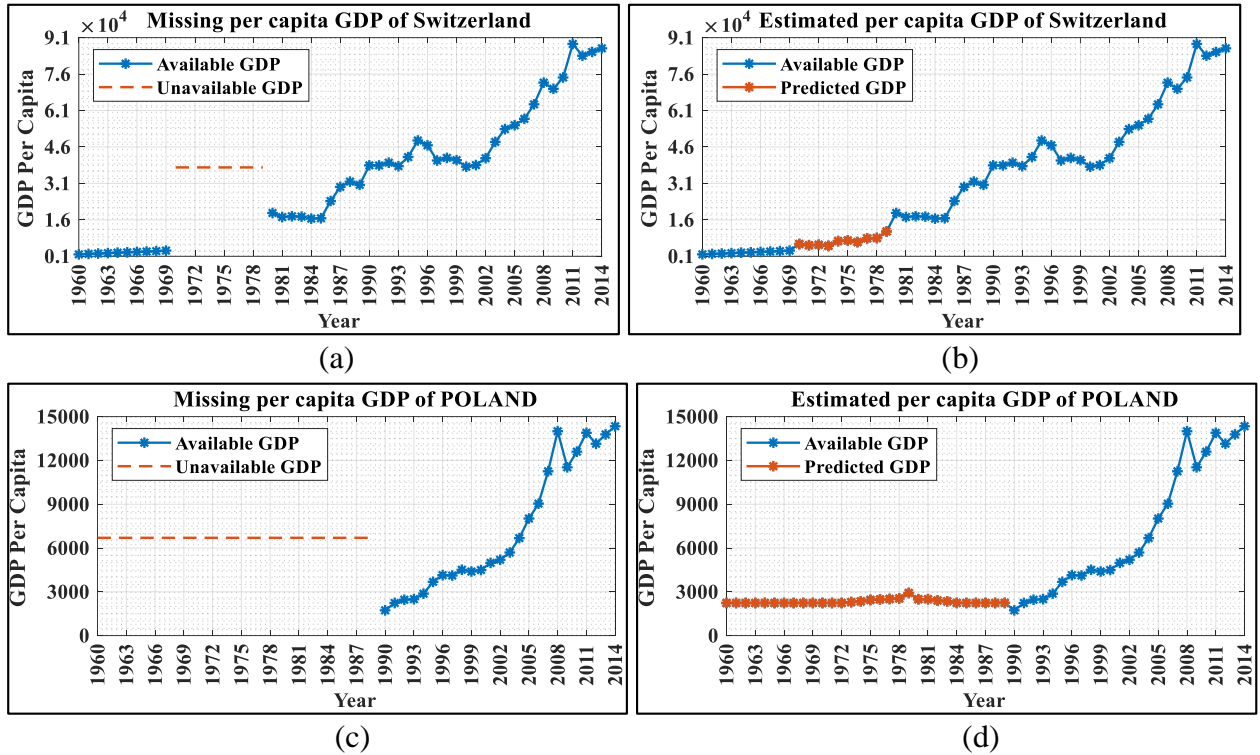

Fig. 15 Developed countries with their missing and estimated per capita GDP

Table 6: Duration of missing per capita GDP values

| Country | Time period of missing values |
|---|---|
| Afghanistan | 1982 to 2000 |
| Iraq | 1965 to 1967, 1991 to 2003 |
| Myanmar | 1960 to 1999 |
| Syria | 2008 to 2014 |
| Yemen | 1960 to 1989 |
| Switzerland | 1970 to 1979 |
| Poland | 1960 to 1989 |

In Machine learning validation data is normally taken between 20% to 40%. However, in this paper we took 33.33% (one third of the available GDP data) as validation data so that a rigorous analysis of the trained model could be done keeping in mind that sufficient data is available for transfer learning during training with other countries data. For better transfer learning, the rest of the available GDP data has been used with the data of IND, CMR, EU and USA for training the SVR, GRNN and ELM. The model which gave the most accurate result on validation dataset was then used to predict the missing GDP values of that country. These results are compiled in Table 7, the first column of which contains the names of the countries whose missing GDP values have been predicted using the model mentioned in the second column. These models predicted the least RMSE error (fourth column) while being trained during DATL-GDP with the source domain depicted in the third column. The fifth column shows the values of the Coefficient of determination.

Per capita GDP values of Afghanistan are missing in the data available at world bank from 1980 to 2000 shown in Table 6 and Fig. 14(a). The first row of the Table 7 depicts that SVR (when trained according to DATL-GDP methodology) by using IND data as source domain produced the least RMSE value of 49.21 on the validation dataset as compared to other SVR, GRNN, or ELM models with different source domains. So, these missing values are then predicted by this SVR model in Fig 14(b).

Iraq is a special case as its per capita GDP values are missing in two different time periods, first from 1965 to 1967 and second from 1991 to 2003 (Fig. 14(c)). These absent values are estimated by ELM in Fig. 14 (d), when CMR is chosen as source domain, as it produced the best result with least RMSE on the validation dataset as compared to the others.

Per capita GDP of Myanmar, unavailable from year 1960 to 1999 (Fig. 14(e)), was estimated by ELM (Fig. 14(f)) according to DATL-GDP approach using EU as source domain, since its RMSE values are the least during validation. No per capita GDP values are available for Syria from year 2008 to 2014 (Fig 14.(g)). For Syria, GRNN predicted smallest RMSE error and similar Coefficient of determination values, when trained using either IND, EU or USA as source domain. To break these ties we randomly chose any one of these countries as source domain, as they had all resulted in similar predictions. So, Syria's unavailable values are predicted in Fig. 14(h) using GRNN with IND source domain. In case of Yemen, values are missing form year 1960 to 1989 (Fig. 14(i)) and these values are then estimated in Fig. 14(j) using GRNN which had the least RMSE on validation data, when trained in accordance with DATL-GDP methodology using CMR as source domain.

The per capita GDP values of Switzerland are not known from year 1970 to 1979 (Fig 15(a)), and these missing values are predicted by ELM in Fig. 15(b) when source domain is EU, as it also predicted the least RMSE when compared to others on the validation dataset. Poland's per capita GDP values are not known from 1960 to 1989 (Fig 15(c)). In case of Poland GRNN predicted the least RMSE error and similar Coefficient of determination in all the cases when source domain was either CMR, IND, EU or USA. As in case of Syria, here too we randomly chose any one of these countries as source domain. So, we chose GRNN trained using USA as source domain to estimate the unknown values and the results are depicted in Fig. 15(d).

Table 7: DATL-GDP Model selection for missing values estimation

| Country | Model | Source domain | RMSE error | Coefficient of determination | RRMSE |
|---|---|---|---|---|---|
| Afghanistan | 'SVR' | 'IND' | 49.21 | 0.48 | 0.24 |
| Iraq | 'ELM' | 'CMR' | 2080.72 | 0.44 | 1.07 |
| Myanmar | 'ELM' | 'EU' | 201.03 | 0.74 | 0.67 |
| Syria | 'GRNN' | 'IND' | 410.94 | 0.51 | 0.54 |
| Yemen | 'GRNN' | 'CMR' | 399.01 | 0.69 | 0.30 |
| Switzerland | 'ELM' | 'EU' | 9462.41 | 0.94 | 0.38 |
| Poland | 'GRNN' | 'USA' | 793.79 | 0.90 | 0.20 |

Except for Poland and Syria, for both of which the lowest RMSE values are same across the three regression methods, models for the rest of the countries resulted in the lowest RMSE values (which are also unique) over validation data, and these most precise models were then chosen for the better estimation of the missing values. In the Table 7, we have also calculated the RRMSE, which is the last column of the table. RRMSE is specifically calculated to gauge the relative error, which is independent of GDP values. For example, let us consider two contrasting scenarios. In the first scenario, suppose the actual per capita GDP of a country in a particular year is very large, say $4,000 and the DATL-GDP model yields a prediction of $4,400. The prediction error then is $400, and relative error is 0.1 (which is very small error). Similarly, in another scenario, let us assume that another country has a very small per capita GDP in a particular year (say, $100) and the DATL-GDP model does a prediction of $500 per capita GDP, then the error is also $400 and the relative error is 4.0 (a huge prediction error).

It can be observed that using relative error in both the scenarios, the first prediction is very accurate, whereas the second prediction is very poor. Therefore, it can be seen from the table that the RMSE error for Afghanistan is 49.21 and for Poland the RMSE error is 793.79; but the prediction of the DATL-GDP model for Poland is more accurate than Afghanistan as Poland's RRMSE (0.20) is smaller than that of the Afghanistan RRMSE (0.24). RRMSE also opens up avenues for future research where one may try to bring down RRMSE value as much as possible for all of the above experiments using more sophisticated techniques like ensemble learning and deep learning etc.

## 7. Conclusion

The rise in global temperatures due to greenhouse gases is causing drastic alterations in the climate. With nearly 99.4% ppb, carbon dioxide is the largest greenhouse gas in the atmosphere. However, studies have proved that $CO_2$ emissions of nations are proportional to their economic might, i.e., $CO_2$ elasticity of GDP is positive and monotonic. Hence, the paper proposes a novel approach to predict per capita GDP of a nation using its $CO_2$ emissions. The proposed approach, which is based on a relatively new concept called transfer learning, is termed as "domain adapted transfer learning for GDP prediction". The efficacy of the proposed approach is demonstrated over the $CO_2$ emission data of four different nations with heterogeneous economic background. We have considered three different regression techniques to present the results comparatively in terms of prediction preciseness. It is also shown that the proposed approach can predict the missing per capita GDP values of a diverse set of countries which were either isolated or war-torn for large amount of time. Existing GDP estimation methods become helpless whenever macroeconomic data of a country is unavailable. However, as the proposed approach does not require macroeconomic data for GDP estimation, it has the edge over other existing GDP estimation methods and potent enough to work as a better alternative in such situations.

### Acknowledgements
Authors are thankful to the anonymous reviewers and the editors for their valuable comments, which have helped us a lot in improving the paper significantly. First author gratefully acknowledges the financial support received from the Department of Science and Technology, Government of India in the form of INSPIRE fellowship. Authors gratefully acknowledge the infrastructural and research facilities provided by the South Asian University, New Delhi through the Computational Intelligence lab of the Department of Computer Science while designing the experiments and conducting investigations.